\documentclass[prd,superscriptaddress,nofootinbib,a4paper,preprintnumbers]{revtex4} 

\usepackage{graphicx}% Include figure files
\usepackage{dcolumn}% Align table columns on decimal point
\usepackage{bm}% bold math
\usepackage{latexsym}
\usepackage{amsfonts}
\usepackage{amssymb}
\usepackage{amsmath}
\usepackage{bbold}
\usepackage{ulem}
\usepackage{url}
\usepackage{xspace}
\usepackage{xcolor}

\allowdisplaybreaks

\newcommand*{\ba}{\begin{eqnarray}}
\newcommand*{\ea}{\end{eqnarray}}
\newcommand*{\mpl}{M_p}
\newcommand{\simgt}{\lower.5ex\hbox{$\; \buildrel > \over \sim \;$}}
\newcommand{\simlt}{\lower.5ex\hbox{$\; \buildrel < \over \sim \;$}}
\newcommand*{\Lag}{{\cal L}}
\newcommand*{\Gdot}{{\dot G}}
\newcommand*{\phid}{\dot{\phi}}

\newcommand*{\p}{\partial}

\newcommand*{\Stuckelberg}{St\"uckelberg~}

\makeatletter
\newcommand{\vast}{\bBigg@{4}}
\newcommand{\Vast}{\bBigg@{5}}
\makeatother

\begin{document}

\title{Massive gravity with non-minimal coupling}

\author{A. Emir G\"umr\"uk\c{c}\"uo\u{g}lu}
\affiliation{Institute of Cosmology and Gravitation, University of Portsmouth\\ Dennis Sciama
Building, Portsmouth PO1 3FX, United Kingdom}

\author{Rampei Kimura}
\affiliation{Waseda Institute for Advanced Study, Waseda University\\
	19th building, 	1-21-1 Nishiwaseda, Shinjuku-ku, Tokyo 169-0051, Japan
}

\author{Kazuya Koyama}
\affiliation{Institute of Cosmology and Gravitation, University of Portsmouth\\ Dennis Sciama
Building, Portsmouth PO1 3FX, United Kingdom}

\date{\today}

\preprint{WU-AP/2001/01}　

\begin{abstract}
We propose new massive gravity theories with $5$ dynamical degrees of freedom. We evade uniqueness theorems regarding the form of the kinetic and potential terms by adopting the ``generalized massive gravity'' framework, where a global translation invariance is broken. By exploiting the rotation symmetry in the field space, we determine two novel classes of theories. The first one is an extension of generalized massive gravity with a non-minimal coupling. On the other hand, the second theory produces a mass term that is 
different from de Rham, Gabadadze, Tolley construction and trivially has $5$ degrees of freedom. Both theories allows for stable cosmological solutions 
without infinite strong coupling, which are free of ghost and gradient instabilities.
\end{abstract}

 \maketitle
 \section{Introduction}
 
Whether a consistent Lorentz-invariant massive gravity theory exists or not has been a long-standing issue for more than seven decades, starting with the pioneering work by Fierz and Pauli \cite{Fierz:1939ix}. The Fierz-Pauli theory of a massive spin-2 field is constructed by choosing a specific combination of the mass terms to have $5$ physical degrees of freedom, but a naive massless limit does not reduce to the massless theory, i.e., linearized general relativity \cite{Zakharov:1970cc,vanDam:1970vg}. This discontinuity was realized to be an artefact of the linear theory and can be solved by taking into account nonlinear completions of the Fierz-Pauli mass term \cite{Vainshtein:1972sx}. Despite this elegant solution, an unwanted sixth ghost degree of freedom called the Boulware-Daser (BD) ghost appears at the nonlinear level \cite{Boulware:1973my}, and thus nonlinear massive gravity was thought to be unstable \cite{Creminelli:2005qk}. Relatively recently, it was shown that the cure to this ghost problem is supplied by adding infinite nonlinear corrections to potential terms \cite{deRham:2010ik,deRham:2010kj}. This ghost-free massive gravity called de Rham-Gabadadze-Tolley (dRGT) theory admits an open-FLRW solution, where the dRGT mass terms exactly behave as the cosmological constant \cite{Gumrukcuoglu:2011ew}, but the scalar and vector kinetic terms of perturbations around this background unfortunately vanishes, which is a signal of a strong coupling problem \cite{Gumrukcuoglu:2011zh} and worse, a non-linear ghost instability \cite{DeFelice:2012mx}. Thus, further extensions of the nonlinear massive gravity should be necessary to accommodate a stable cosmological solution. 
 
Although one can simply extend the dRGT theory by introducing other fields such as the quasi-dilaton theory \cite{DAmico:2012hia}, mass-varying massive gravity \cite{Huang:2012pe}, and Hassan-Rosen bigravity \cite{Hassan:2011zd}, it might be more interesting if the massive gravity can be further generalized without invoking additional degrees of freedom. So far, such extensions have been intensively investigated with the various approaches, but none of them are successful at this point \cite{Folkerts:2011ev,Hinterbichler:2013eza,Kimura:2013ika,deRham:2013tfa,Gao:2014jja}. 
All of the above investigations rely on theories invariant under the Poincar{\'e} symmetry in the internal field space consisted of the \Stuckelberg fields $\phi^a$, which are responsible for restoring the general covariance \cite{ArkaniHamed:2002sp}.
However, once we abandon the translation invariance while keeping the global Lorentz invariance, a natural extension of the dRGT theory can be accessible \cite{deRham:2014lqa}. In this theory called the generalized dRGT theory, the constant parameters in the graviton potential are promoted to be arbitrary functions of four auxiliary fields $\phi^a$, and the total number of physical degrees of freedom remains the same as the dRGT theory, i.e., the BD ghost is absent. Furthermore, it has been recently shown that all perturbations around the open-FLRW background are free from any instabilities \cite{Kenna-Allison:2019tbu}. 
  
Now, it is interesting to know how far we can generalize (global) Lorentz-invariant massive gravity theories. In any massive gravity theory with five degrees of freedom, there exists the Hamiltonian constraint in unitary gauge, and this guarantees the absence of the BD ghost \cite{Hassan:2011tf}. In the \Stuckelberg language, the degeneracy of the kinetic matrix of four scalar fields $\phi^a$ leads one of them to be non-dynamical, implying that the would-be BD ghost is successfully eliminated \cite{deRham:2011rn}. Therefore, imposing the degeneracy of the four scalar fields $\phi^a$, one should be able to explore new theories of massive gravity. In the present paper, we investigate the possibility of extending massive gravity theory which preserves the global Lorentz symmetry and derive two distinct classes of ghost-free theories. 

To derive these novel theories, we make use of two specific properties of these field theories that become available once translation invariance is dropped. First, introducing the matter fields after performing conformal transformation of the physical metric generates a non-minimal coupling of the \Stuckelberg fields to the curvature. Second, we show that conformal and disformal deformations of the fiducial metric may evade the reappearance of the BD mode under certain conditions, which only leave the two unconnected theory classes that we present here. The first theory is a direct extension of generalized massive gravity with modified fiducial metric and non-minimal coupling. The derivation of this theory also reveals how one can generate the original generalized massive gravity action starting from constant parameter dRGT theory. The second theory is different from dRGT type constructions by directly projecting out one degree of freedom and is thus closely linked to Lorentz breaking theories. 

This paper is organized as follows. In Sec.~\ref{sec:tbreaking}, we briefly review the dRGT massive gravity and discuss the implications of breaking the translation invariance, which allows a conformal transformation of the physical metric. In Sec.~\ref{sec:disformal}, we generalize the dRGT mass terms by using disformal transformation acted on the fiducial metric and summarize necessary conditions to eliminate the would-be BD ghost. In Sec.~\ref{sec:3+1}, we investigate a kinetic Lagrangian including a non-minimal coupling and derive a degeneracy condition by using $3+1$ decomposition. In Sec.~\ref{sec:cosmology}, we derive background equations and quadratic actions for perturbations in the two theories obtained in Sec.~\ref{sec:3+1} and show that all perturbations around FLRW background are free of ghost and gradient instabilities. Sec.~\ref{sec:summary} is devoted to summary.

\section{Breaking translation invariance and non-minimal coupling}
\label{sec:tbreaking}
In this section, we briefly introduce the ghost-free massive gravity and argue the existence of a non-minimal coupling in a global Lorentz invariant massive gravity. Let us start with the ghost-free dRGT massive gravity, which is given by \cite{deRham:2010ik,deRham:2010kj}
\ba
	S_{\rm dRGT} = \int d^4x\,\sqrt{-{g}}\frac{\mpl^2}{2}\left[{R} -2m^2\sum_{n=0}^3\,{\beta}_n \,e_n(\sqrt{{g}^{-1}f})\right] + S_{\rm m}  [g, \psi] \,,
	\label{dRGT}
\ea
where $\beta_n$ are constant parameters,
$S_{\rm m}  $  is the matter action, 
and the dRGT potential terms are built out of $e_n $, 
\ba
e_0 ({\cal Q})
 &=& 1\,, \\
e_1 ({\cal Q})&=& [{\cal Q}]\,, \\
e_2 ({\cal Q})&=& {1 \over 2!} \left([{\cal Q}]^2-[{\cal Q}^2]\right)\,, \\
e_3 ({\cal Q})&=& {1 \over 3!} \left([{\cal Q}]^3-3[{\cal Q}][{\cal Q}^2]
+2[{\cal Q}^3] ) \right)\,, \\
e_4 ({\cal Q})&=& {1 \over 4!} \left(
[{\cal Q}]^4-6[{\cal Q}]^2[{\cal Q}^2]
+3[{\cal Q}^2]^2+8[{\cal Q}][{\cal Q}^3]-6[{\cal Q}^4]\right)\,.
\label{eq:elementarypoly}
\ea
Here, $[{\cal Q}]$ denotes the trace of the matrix ${\cal Q}$, 
a square root of the matrix represents the matrix satisfies 
$\sqrt{{\cal Q}}^\mu_{~\rho}\sqrt{{\cal Q}}^\rho_{~\nu} = {\cal Q}^{\mu}_{~\nu}$,
and $f_{\mu\nu}$ is the fiducial metric, which is defined through
\begin{equation}
	f_{\mu\nu} \equiv \eta_{ab}\,\partial_\mu \phi^a\,\partial_\nu \phi^b\,,
	\label{fiducialdRGT}
\end{equation}
and $\eta_{ab}$ is the Minkowski metric with $a, b = 0, 1, 2, 3$.  
The action \eqref{dRGT} is manifestly invariant under the 
 Poincar{\'e} symmetry in the internal field space. Once we abandon the global translation invariance $\phi^a \to \phi^a + c$, the scalar function $X=\eta_{ab}\phi^a\phi^b=\phi^a\phi_a$ can promote the constant parameters $\beta_n$ to be functions of $X$, and the resultant theory, i.e. the generalized massive gravity, still enjoys a global Lorentz invariance \cite{deRham:2014lqa}. 
In the present paper, we seek further extensions to such a global Lorentz invariant massive gravity. 

Let us first consider the conformal transformation utilizing the scalar $X$,
\ba
{\tilde g}_{\mu\nu} = G(X) g_{\mu\nu} \,.
\ea
Performing this transformation to the action \eqref{dRGT}, the gravitational part of the Lagrangian becomes
\ba
	S &=& \int d^4x\,\sqrt{-\tilde{g}}\left[\frac{\mpl^2}{2}\,R[{\tilde g}] -\mpl^2m^2\sum_{n=0}^3\,\beta_n \,e_n(\sqrt{\tilde{g}^{-1}f})\right] \nonumber\\
	&=& \int d^4x\,\sqrt{-g}\left[ 
	\frac{\mpl^2 }{2} G\left(R+\frac{3}{2}\,\nabla^\mu \log G\,\nabla_\mu \log G\right) -\mpl^2m^2\sum_{n=0}^3 {\tilde \beta}_n (X)e_n(\sqrt{g^{-1}f})\right]\,,
\ea
where we defined  the rescaled parameters as
\begin{equation}
	{\tilde \beta}_n (X)\equiv \beta_n \,G(X)^{\frac{4-n}{2}} \,.
\end{equation}
The kinetic part of the Lagrangian contains 
the non-minimal coupling with the conformal factor and its counter term, 
which can be rewritten as
\ba
\frac{3}{2}\,\nabla^\mu \log G(X)\,\nabla_\mu \log G(X) =  {6 G_X^2 \over G^2} \phi_a \phi_b g^{\mu\nu}\partial_\mu \phi^a \partial_\nu \phi^b \,.
\label{nonminimal}
\ea
Thus, once the translation invariance is broken, the Einstein-Hilbert term can be non-minimally coupled with $\phi^a$ with a suitable counter term. 
Furthermore, the parameters ${\tilde \beta}_n$ after the transformation are the arbitrary functions of $X$, and this is reminiscent of the generalized massive gravity. 
The conformal scaling in the mass terms can be interpreted as a redefinition of the fiducial metric,  
${\tilde g}^{-1}f \to g^{-1} (G^{-1} f) $.
This implies that the fiducial metric no longer needs to be of the form \eqref{fiducialdRGT},  and one can deform the fiducial metric by appropriately contracting the Lorentz indices by $\eta_{ab}$ and $\phi_a$ and introducing arbitrary functions of $X$. 

\section{Disformal deformations of the fiducial metric}
\label{sec:disformal}

As we have seen in the previous section, a conformal transformation of the physical metric would lead us to a new extension of massive gravity including a non-minimal coupling. Another effect of the conformal transformation is to rescale the fiducial metric by a conformal factor that depends on the \Stuckelberg fields. 
In this section we explore this option further by starting with the most general deformation of the fiducial metric,
\ba
{\tilde f}_{\mu\nu,I} = L_{ab,I}\,\partial_\mu \phi^a\,\partial_\nu \phi^b\,,
\label{eq:disformalreference}
\ea
with 
\ba
L_{ab,I} = C_I (X) \eta_{ab} + D_I (X) \phi_a \phi_b \,,
\ea
where $I$ is a label that will be assigned to each mass term. We then construct the following square-root matrix
\ba
Q^\mu_{~\nu, I} \equiv \left(\sqrt{g^{-1}{\tilde f}_I} \right)^\mu_{~\nu} \,.
\label{QI}
\ea 
Here, if all $C_I=1$ and $D_I=0$, this reduces to the square-root matrix in dRGT massive gravity.
Then, we consider the action 
with the Einstein-Hilbert term
and mass terms, 
\begin{equation}
S = \int d^4x \sqrt{-g} \frac{\mpl^2 }{2} \Bigl[R[g]
-2\,m^2 {\cal L}_{\rm mass}
\Bigr]\,,
\label{action:quadratic}
\end{equation}
where 
\ba
{\cal L}_{\rm mass} &=&	
\beta(X) \,[Q_{\beta}]+ 
\gamma_1(X)\,[Q_{\gamma_1}]^2+\gamma_2(X)\,[Q_{\gamma_2}^2] 
+ \delta_1(X)\,[Q_{\delta_1}]^3 + \delta_2(X)\,[Q_{\delta_2}][Q_{\delta_2}^2]+ \delta_3(X)\,[Q_{\delta_3}^3] 
\nonumber\\
&&+\sigma_1(X)\,[Q_{\sigma_1}]^4
+\sigma_2(X)\,[Q_{\sigma_2}]^2\,[Q_{\sigma_2}^2]
+\sigma_3(X)\,[Q_{\sigma_3}^2]^2
+\sigma_4(X)\,[Q_{\sigma_4}]\,[Q_{\sigma_4}^3]
+\sigma_5(X)\,[Q_{\sigma_5}^4]\,.
\label{eq:Lmass-deformed}
\ea  

In general, the four scalar components in the \Stuckelberg fields $\phi^a$ bring an extra degree of freedom, i.e. the BD ghost. To avoid this, the action should be arranged such that one of the component has no dynamics. This can be actually realized by the degeneracy of the kinetic matrix of the \Stuckelberg fields \cite{deRham:2011rn}. Then, it generates a primary and  subsequent constraints, and the BD ghost can be successfully eliminated. We here adopt this approach to obtain BD ghost-free conditions rather than the decoupling limit approach \cite{deRham:2010ik}. 

The majority of the non-perturbative proofs of the absence of ghost in dRGT theory relies on the unitary gauge \footnote{When translation invariance is broken, the unitary gauge, $\phi^a = \delta^a_{\mu} x^{\mu}$, is no longer a convenient choice. Instead, the gauge fixing with $\phi^0 = \sqrt{t^2 + x^ix_i}$ and $\phi^i = x^i$ so that $\phi^a\phi_a = -t^2$ would be convenient as introduced in \cite{deRham:2014lqa}.}. 
There are several non-perturbative proofs of the absence of the BD ghost in the presence of the St\"uckelberg fields \cite{deRham:2011rn,Kluson:2011rt,Hassan:2012qv,Kluson:2012gz,KLUSO?:2013spa,Kugo:2014hja} but these rely on specific techniques not applicable to the case at hand. 
Due to the square-root form of the building block tensor \eqref{QI} it is highly challenging to derive the degeneracy conditions exactly. However, around fixed backgrounds we can obtain necessary (but not sufficient) degeneracy conditions as follows \footnote{These conditions were derived by combining the degeneracy conditions around Bianchi type V and a static non-diagonal metric with a preferred direction. The details of the calculation can be found in Appendix \ref{app:fixedbg-degeneracy}.}. 
In the mass term, we introduced $20$ arbitrary functions to represent the $10$ disformal transformations. The following conditions reduce the number of independent functions down to $4$ 
\begin{align}
\frac{D_{\beta}}{C_{\beta}}
	&=\frac{D_{\gamma_1}}{C_{\gamma_1}}=\frac{D_{\gamma_2}}{C_{\gamma_1}}= \frac{D_{\delta_1}}{C_{\delta_1}}
= \frac{D_{\delta_2}}{C_{\delta_1}}
= \frac{D_{\delta_3}}{C_{\delta_1}}
= \frac{D_{\sigma_1}}{C_{\sigma_1}}
= \frac{D_{\sigma_2}}{C_{\sigma_1}}
= \frac{D_{\sigma_3}}{C_{\sigma_1}}
= \frac{D_{\sigma_4}}{C_{\sigma_1}}
= \frac{D_{\sigma_5}}{C_{\sigma_1}}\,,\nonumber\\
C_{\gamma_1}&=C_{\gamma_2}\,,\nonumber\\
C_{\delta_1}&=C_{\delta_2} =C_{\delta_3}\,,\nonumber\\
C_{\sigma_1}&=C_{\sigma_2} =C_{\sigma_3}=C_{\sigma_4}=C_{\sigma_5}\,,
\label{eq:condition-transformation-functions}
\end{align}
which implies that all mass terms are built out of a universal field space metric $\eta_{ab} + \frac{D}{C}\,\phi_a\phi_b$ with three independent conformal factors for each orders. Using these relations, the conditions that the mass functions need to satisfy can be written as
\begin{align}
(C_{\gamma_1}+X\,D_{\gamma_1})(\gamma_1+\gamma_2) &=0\,,\nonumber\\
	(C_{\delta_1}+XD_{\delta_1})^{3/2}(
	\delta_1+\delta_2+\delta_3) &=0\,,\nonumber\\
\sqrt{C_{\delta_1}}(C_{\delta_1}+XD_{\delta_1})(3\,\delta_1
+\delta_2 )&=0\,,\nonumber\\
(C_{\sigma_1}+XD_{\sigma_1})^2(\sigma_1+\sigma_2+\sigma_3+\sigma_4+\sigma_5) &=0\,,\nonumber\\
\sqrt{C_{\sigma_1}}(C_{\sigma_1}+XD_{\sigma_1})^{3/2}(
4\,\sigma_1	+2\,\sigma_2+\sigma_4) &=0\,,\nonumber\\
C_{\sigma_1}(C_{\sigma_1}+XD_{\sigma_1})( 6\,\sigma_1+\sigma_2) &=0
	\,,\nonumber\\
C_{\sigma_1}(C_{\sigma_1}+XD_{\sigma_1})( 3\,\sigma_1+\sigma_2+\sigma_3) & = 0 \,.
\label{eq:condition-mass-functions}
\end{align}
There are two ways to satisfy these conditions. For $(1+D_I\,X/C_I)\neq0$, the first solution is 
\begin{equation}
\gamma_1=-\gamma_2\,,\qquad
\delta_1=-\frac{\delta_2}{3}= \frac{\delta_3}{2}\,,\qquad
\sigma_1=-\frac{\sigma_2}{6}=\frac{\sigma_3}{3}=\frac{\sigma_4}{8}=-\frac{\sigma_5}{6}\,,
\label{DC1}
\end{equation}
which simply is the dRGT tuning with mass term
\footnote{With this tuning, the quartic mass term $\sqrt{-g}\,\beta_4(X)e_4(Q) = \sqrt{-\tilde{f}}\,\beta_4(X)$ becomes a boundary term, similarly to constant mass dRGT.}
\begin{align}
{\cal L}_{\rm mass} =&\beta_1(X)\,e_1(Q)+	\beta_2(X)\,e_2(Q) + \beta_3(X)\,e_3(Q)+\beta_4(X)\,e_4(Q)\,,
\end{align}
where $Q \equiv \eta_{ab} + D(X)\phi_a\phi_b$ and by absorbing the conformal factors at each order into the mass function, we reduce the number of arbitrary functions down to $4$ with the following definitions
\begin{equation}
\beta_1 \equiv \,C_{\beta}\,\beta\,,\qquad
\beta_2 \equiv 2\,C_{\gamma_1}\,\gamma_1\,,\qquad
\beta_3 \equiv 3!\,C_{\delta_1}^{3/2}\delta_1\,,\qquad
\beta_4  \equiv 4!\,C_{\sigma_1}^{3/2}\,\sigma_1\,,\qquad
D  \equiv \frac{D_{\gamma_1}}{C_{\gamma_1}}\,.
\end{equation}
In other words, there can only be one field space metric, disformally related to the original Minkowski metric. The case with a field space-metric that can explicitly depend on the $\phi^a$ was already argued to be ghost-free in \cite{Hassan:2012qv}. Thus, starting from  dRGT theory \eqref{dRGT} with constant mass parameters, one can consider deformations of the field space metric with different conformal coefficients at different orders and generate the generalized massive gravity action.

The second option for satisfying Eq.\eqref{eq:condition-mass-functions} is to fix the disformal term in the field space metric as
\begin{equation}
\frac{D_{\gamma_1}}{C_{\gamma_1}} = -\frac{1}{X}\,,
\label{DC2}
\end{equation}
which leaves all of the mass functions unconstrained.
This condition is equivalent to having a field space metric proportional to a projection tensor  $P_{ab}$ defined as 
\ba
P_{ab} \equiv \left( \eta_{ab} - \frac{\phi_a\phi_b}{X}\right) \,,
\ea
which projects onto surfaces in the field space defined by normal vector $\phi^a$, while the conformal factors can be absorbed in the individual mass functions. This precise combination guarantees that the derivative of one of the directions is absent in $\tilde{f}_{\mu\nu}$. The mass term constructed with this projection tensor cannot be combined with (unprojected) dRGT-type terms where the degeneracy disagrees with the one imposed by the projection.
As we will see in the next section, the projected mass terms naturally lack the BD mode, and are reminiscent of some of the Lorentz violating massive gravity theories \cite{Lin:2013aha,Lin:2013sja}. The difference however is that the time direction in our theory remains unspecified, thanks to the explicit dependence on $\phi^a$ allowed by the broken translation invariance.

\section{Evading BD ghost in 3+1 decomposition}
\label{sec:3+1}
In this section, we extend the previous analysis to include a non-minimal coupling. To see the kinetic structure of the non-minimal coupling, we consider the following action 
\ba
S
&=&
\int {\rm d}^4 x \sqrt{-g} \frac{\mpl^2}{2}\Biggl[ G(X) R[g]+ F(X) [Y] +  A (X) [W]
\Biggr]\,,
\label{action:3+1}
\ea
where we defined $W$ and $Y$ as
\ba
W^\mu_{~\nu} = (g^{-1}f)^\mu_{~\nu} \,, \qquad 
Y^\mu_{~\nu} \equiv g^{\mu\nu} \phi_a  \phi_b \partial_\mu \phi^a\partial_\nu\phi^b \,.
\label{eq:defWY}
\ea
These two terms are responsible for the degeneracy of the kinetic matrix with the non-minimal coupling $G(X)R$. Other higher order candidates such as $[W^2]$ and $[Y^2]$ as well as the terms involving the square-root tensor cannot be counter terms for the non-minimal coupling. This is because the non-minimal coupling term is quadratic in the extrinsic curvature and contains a linear mixing between $\dot{\phi}_a$ and the extrinsic curvature (see \eqref{kinetic} below). Thus the counter terms need to be quadratic in $\dot{\phi}_a$. The $F$ term is the one that arises from the conformal transformation of the physical metric \eqref{nonminimal}. $F$ and $A$ terms can be included in $\gamma_2$ term of \eqref{action:quadratic}, and these terms can be generated by the conformal and disformal transformation of the fiducial metric.

We investigate the degeneracy using a $3+1$ decomposition. This is similar to the analysis performed to find degenerate higher order scalar tensor theories \cite{Langlois:2015cwa,Crisostomi:2016czh, BenAchour:2016fzp}, but the difference is that we are interested in finding degeneracies between metric variables and four scalar fields $\phi^a$ in our case.

Let us define normal vector $n^{\mu}$ of each spacelike hypersurface $\Sigma_t$, which satisfies $n^{\mu}n_{\mu} = -1$. The induced metric $\gamma_{\mu\nu}$ on a spacelike hypersurface $\Sigma_t$ is then defined by 
$\gamma_{\mu\nu} = g_{\mu\nu} + n_{\mu} n_{\nu}$.
The derivative of the normal vector can be expressed as  $\nabla_{\mu} n_{\nu} = - n_{\mu} a_{\nu} + K_{\mu\nu} $,
where $a_\mu$ is the acceleration vector, $a^{\mu} := n^{\nu}\nabla_{\nu}n^{\mu}$, 
and $K_{\mu\nu}$ is the extrinsic curvature,
$K_{\mu\nu} := \gamma_{\mu}{}^{\rho} \gamma_{\nu}{}^{\sigma}\nabla_{\rho}n_{\sigma}$.
By using the normal vector and the induced metric, the partial derivative of four scalar fields can be decomposed as
\begin{align}
\partial_\mu \phi^a &= - n_{\mu} \phid^a  + D_{\mu}\phi^a,
\end{align}
where we defined $\phid^a := n^{\mu}\partial_{\mu} \phi^a$ and $D_{\mu}\phi^a := \gamma_{\mu}{}^{\nu} \partial_{\nu} \phi^a$.

Then the kinetic part of the Lagrangian \eqref{action:3+1} can be expressed as  
\ba
\Lag_{\rm kin}=
{\cal A}^{ab} \phid_a \phid_b + {\cal C}^{a\mu\nu} \phid_a K_{\mu\nu} +{\cal F}^{\mu\nu\rho\sigma} K_{\mu\nu} K_{\rho\sigma},
	\label{kinetic}
\ea
with
\begin{subequations}
	\begin{align}
	{\cal A}^{ab} =&  - A \eta^{ab} - F \phi^a \phi^b \,, \qquad
	{\cal C}^{a\mu\nu} =- 4\eta^{ab} G_X \phi_b \gamma^{\mu\nu} \,, \qquad
	{\cal F}^{\mu\nu\rho\sigma}= 
	G ( \gamma^{\mu(\rho}\gamma^{\sigma)\nu} - \gamma^{\mu\nu}\gamma^{\rho\sigma}),
	\end{align}
\end{subequations}
where we have used $R = {}^{(3)}R + K_{\mu\nu}K^{\mu\nu}-K^2 -2 \nabla_{\mu}(a^\mu -K n^{\mu} )$,
and ${}^{(3)}R$ stands for the three-dimensional Ricci scalar composed by the spatial metric, $\gamma_{\mu \nu}$. 
Now the canonical momenta are given by 
\ba
\pi^a &=& {\delta {\cal L} \over \delta \phid_a} = 2 {\cal A}^{ab} \phid_b + {\cal C}^{a\mu\nu} K_{\mu\nu} +  ({\rm no~time~derivative~terms}) \,,\\
\pi^{\mu\nu} &=& {\delta {\cal L} \over \delta K_{\mu\nu}} = {\cal C}^{a\mu\nu} \phid_a + 2 {\cal F}^{\mu\nu\rho\sigma} K_{\rho\sigma} +  ({\rm no~time~derivative~terms}) \,.
\ea
The existence of a primary constraint is ensured if and only if the linear combination of the canonical momenta 
\ba
\Psi \equiv \, \alpha_1 \phi^a \pi_a + \alpha_2 \gamma^{\mu\nu} \pi_{\mu\nu} \,,
\label{DCpre}
\ea
where $\alpha_1$ and $\alpha_2$ are constants,
is independent of the velocities $K_{\mu\nu}$ and $\phid^a$, i.e., 
\ba
\frac{\partial \Psi}{\partial (\phi_a\phid^a)} = 0\,, \qquad 
\frac{\partial \Psi}{\partial K} = 0 \,.
\ea
These equations provide the degeneracy condition, 
\ba
G A + G F X -6 G_X^2 X =0 \,.
\label{DC}
\ea
As long as this condition is satisfied, the primary (and subsequent) constraint should exist, and the absence of the BD ghost is ensured. 
One should note that this agrees with the condition \eqref{eq:bianchi5conditions} in Appendix \ref{app:fixedbg-degeneracy}, in the absence of $\gamma_1$ and $\gamma_2$.

We now discuss how to include the mass terms obtained in the previous section to this construction. We start with the first option \eqref{DC1}. 
After solving the degeneracy condition \eqref{DC} in terms of $F$, the kinetic Lagrangian can be written as
\ba
\Lag_{\rm kin}=
{\cal F}^{\mu\nu\rho\sigma} \left(K_{\mu\nu}+ {G_X \over G}\phi_a {\dot \phi}^a \gamma_{\mu\nu}\right) \left(K_{\rho\sigma}+ {G_X \over G}\phi_b {\dot \phi}^b \gamma_{\rho\sigma}\right)
- A \phid_a \phid^a \,.
\ea 
Therefore, in the absence of $A$, the $F$ term can be absorbed into the extrinsic curvature by a field redefinition, and we can safely add the mass terms of the generalized massive gravity with disformal field space metric, as discussed in the previous section. 
On the other hand, in the presence of $A$ term, the degeneracy condition is not compatible with the dRGT potential terms since $A$ term can be included in $\gamma_2$ in \eqref{action:quadratic} and it changes the ratio $D/C$. Therefore, $A=0$ should be required under the condition $\gamma_1 + \gamma_2=0$ to ensure the absence of the BD ghost.
Therefore, the ghost-free extension of generalized massive gravity with the non-minimal coupling is given by
\ba
S = \int d^4x\,\sqrt{-{g}}\frac{\mpl^2}{2} \left[ 
GR[g]+ 
 {6 G_X^2 \over G} [Y]
 -2m^2\sum_{n=0}^3\,{\beta}_n(X) \,e_n(\sqrt{{g}^{-1}\tilde{f}})\right] + S_{\rm m}  [g, \psi] \,.
\label{action: NMMG}
\ea

Let us now discuss another option \eqref{DC2} obtained in the previous section. As mentioned in the previous section, this is nothing but the projection onto the $\phi^a$ direction. With this projection operator, $P_{ab} = \eta_{ab} - {\phi_a\phi_b /X}$, one can construct the following fiducial metric ${\bar f}_{\mu\nu}$ and the building block tensor $Z^{\mu}_{\;\;\nu}$:
\ba
{\bar f}_{\mu\nu} =P_{ab}\partial_\mu \phi^a \partial_\nu \phi^b \,, \qquad 
Z^\mu_{~\nu}= (g^{-1}{\bar f})^\mu_{~\nu}\,.
\ea
With the trace of this tensor $Z^\mu_{~\nu}$ and the degeneracy condition \eqref{DC}, we can rewrite the Lagrangian as 
\ba
G \, R + F[Y] +  A [W] = G \,R  + {6 G_X^2 \over G} [Y] + A [Z] \,,
\ea 
where we have used the relation $Z = W - Y/X$. The potential term $[Z]$ can be further generalized, and 
we arrive at ghost-free massive gravity, which is distinct from \eqref{action: NMMG}, 
\ba
S=
\int {\rm d}^4 x \sqrt{-g} \frac{\mpl^2}{2}\Biggl[ G \,R  + {6 G_X^2 \over G} [Y] 
+ m^2  \,{\cal U} \bigl(X, [Z], [Z^2], [Z^3] \bigr)
\Biggr] + S_{\rm m}  [g, \psi] \,,
\label{action:NMP}
\ea
where $ {\cal U} $ is now an arbitrary function of $X, [Z], [Z^2]$ and $[Z^3]$. Note that higher order terms $[Z^n]$ with $n \geq 4$ can be also included in $ {\cal U} $, but they can be always reduced to lower order terms by Cayley-Hamilton theorem.  
The absence of the BD ghost can be simply understood as follows. The contribution to the momenta $\pi^a$ from the potential ${\cal U} $ is proportional to the projection tensor $P_{ab}$, and thus ${\cal U} $ does not contribute to  \eqref{DCpre} automatically. 
Thus, it is manifest that the potential ${\cal U}$ including higher order is still compatible with the condition \eqref{DCpre}, and the action \eqref{action:NMP} is therefore ghost-free.
In the Appendix~\ref{app:NMP}, we show an explicit derivation of \eqref{action:NMP} starting with the most general mass terms up to quadratic order composed of $W$ and $Y$. 
Finally, let us mention the relation between the action with \eqref{DC2} in the previous section and the action \eqref{action:NMP}. 
The trace of any square-root tensor can be written in terms of one without square-roots (see Ref.~\cite{Gumrukcuoglu:2019rsw} for details). This allows us to rewrite \eqref{action:quadratic} 
in terms of $[Z^n]$, and the theory \eqref{action:quadratic}  with \eqref{DC2} is manifestly included in \eqref{action:NMP}.

\section{Cosmological background and perturbations}
\label{sec:cosmology}

In the previous section, we have proposed two distinct types of the extended theories \eqref{action: NMMG} and \eqref{action:NMP}. It is now interesting to ask whether all the polarization modes are healthy or not on physically relevant backgrounds. To this end, we study FLRW cosmologies in the obtained theories \eqref{action: NMMG} and \eqref{action:NMP} and derive the conditions for avoiding the ghost and gradient instabilities. Following \cite{Gumrukcuoglu:2011ew}, we choose the St\"uckelberg fields as\footnote{The requirement of homogeneity and isotropy constrains the St\"uckelberg configuration uniquely. However, if one relaxes the requirement, it is possible to realize backgrounds that are approximately FLRW within the observable patch \cite{DAmico:2011eto}, which allows more general configuration of the scalar fields.}
\begin{equation}
\phi^0 = f(t)\,\sqrt{1+\kappa(x^2+y^2+z^2)}\,,
\qquad
\phi^i = f(t)\,\sqrt{\kappa}\,x^i\,,
\label{eq:Stuckelberg}
\end{equation}
and the physical metric is chosen to be compatible with this choice, i.e. an open FLRW metric. We remark that due to the explicit dependence on $\phi^a\phi_a$ in the action, the requirement of homogeneity allows only for an open universe solution.
Including perturbations, the physical metric is given by
\ba
ds^2 = -(1 + 2 \Phi ) dt^2 + a(t) (\partial_i B + B_i ) dt dx^i + a(t)^2
(\Omega_{ij} + h_{ij}) \,,
\ea
where $\Omega_{ij}$ is the spatial metric with a constant curvature $-\kappa<0$
\begin{equation}
\Omega_{ij} = \delta_{ij}-\frac{\kappa\,x^ix^j}{1+\kappa\,x^kx^k}\,,
\end{equation}
and we decompose spatial components of metric perturbations as follows:
\ba
h_{ij} = 2\psi \Omega_{ij} +\left(D_iD_j - {1 \over 3} \Omega_{ij} D_k D^k\right) E + {1 \over 2} (D_i E_j + D_j E_i) + \gamma_{ij} \,. 
\ea
Here, $D_i$ is the covariant derivative compatible with the $\Omega_{ij}$ metric. The vector perturbations are divergence-free $D^i E_i = D^i B_i =0$, and the tensor perturbation is  divergence and trace-free $D^i \gamma_{ij} = \Omega^{ij} \gamma_{ij} = 0$. 
For the four scalar fields $\phi^a$, we fix the gauge such that $\phi^a = \langle \phi^a \rangle$, i.e., no perturbation.
As for the matter sector, we introduce a k-essence field to mimic an irrotational perfect fluid: 
\begin{equation}
S_{\rm m}  = \int d^4 x \sqrt{-g} \, P(\Theta)\,,\qquad
\Theta\equiv -\frac12\,\partial_\mu \chi\,\partial^\nu\chi\,,
\end{equation}
where $\chi$ is a scalar field, which can be decomposed into the background $\chi_0$ and perturbation $\delta\chi$ as
\begin{equation}
\chi = \chi_0 + \mpl \,\delta \chi\,.
\label{eq:matterperturbations}
\end{equation}
Throughout this section, we use the equivalent energy density, pressure, and the sound speed of the k-essence field, defined as
\ba
\rho_\chi = 2 P_{\Theta} \Theta - P \,, \qquad
p_\chi = P \,, \qquad
c_\chi^2 = \frac{P_{\Theta}}{2P_{\Theta\Theta}\Theta + P_{\Theta}} \,,
\ea
where $P_\Theta = \partial P / \partial \Theta$ and $P_{\Theta\Theta} = \partial^2 P / \partial \Theta^2$. 
Then the equation of motion for $\chi_0$ is given by
\ba
\dot{\rho}_\chi + 3H (\rho_\chi + p_\chi) &=&0 \,.
\label{EOMchi}
\ea
Hereafter, we also use
\ba
\xi \equiv \frac{\sqrt{\kappa }f}{a} \,,
\ea
in replacement of $f$. In deriving second order actions in each sector, we expand tensor, vector, and scalar perturbations in terms of harmonics, for example,
\ba
\gamma_{ij} &=& \int k^2 dk \gamma_{|\vec{k}|} Y_{ij} (\vec{k}, \vec{x}) \,, \\
B_{i} &=& \int k^2 dk B_{V,|\vec{k}|} Y_{i} (\vec{k}, \vec{x}) \,, \\
\Phi &=& \int k^2 dk \Phi_{S, |\vec{k}|} Y (\vec{k}, \vec{x}) \,,
\ea
with $D_l D^l Y_{ij} = -k^2 Y_{ij}$ and $D^i Y_{ij} = \Omega^{ij} Y_{ij}=0$ for tensor harmonics, 
 $D_l D^l Y_{i} = -k^2 Y_{i}$ and $D^i Y_{i}=0$ for vector harmonics,  and $D_l D^l Y = -k^2 Y$  for scalar harmonics.
Other perturbations,  $E_i$, $B$, $\psi$, and $E$ are defined in a similar manner.

\subsection{Non-minimal generalized massive gravity}
In this subsection, we focus on the non-minimally coupled generalized massive gravity \eqref{action: NMMG}, with general reference metric
\begin{align}
S = \int d^4x\,\sqrt{-{g}}\frac{\mpl^2}{2} \left[ 
G\,R[g]+ 
 {6 G_X^2 \over G} [Y]
 -2m^2\sum_{n=0}^3\,{\beta}_n(X) \,e_n(\sqrt{{g}^{-1}\tilde{f}})\right] + S_{\rm m}  [g, \chi] \,,
\label{action:NMMG-re}
 \end{align}
where $Y$ is defined in Eq.\eqref{eq:defWY} and the disformal fiducial metric is
\begin{equation}
\tilde{f}_{\mu\nu} = \left(\eta_{ab} + D\,\phi_a\phi_b\right)\partial_\mu\phi^a\partial_\nu\phi^b\,.
\end{equation}
For the field configuration \eqref{eq:Stuckelberg}, the background line element for the fiducial metric is given by
\begin{equation}
\tilde{f}_{\mu\nu} dx^\mu dx^\nu = -\dot{f}^2(1-f^2\,D)\,dt^2+\kappa\,f^2\,dx^i \Omega_{ij}dx^j\,,
\end{equation}
so the disformal part of the field space metric $D$ shifts the lapse function of the fiducial metric to $\dot{f}\to\dot{f}\sqrt{1-f^2\,D}$, while the scale factor in the fiducial metric continues to be $\sqrt{\kappa}\,f$. 

\subsubsection{Background equations}
We define the following functions which will be useful later on
\begin{align}
\rho_{m,g} &\equiv\beta_0 +3\,\xi\,\beta_1 +3\,\xi^2\beta_2+\xi^3\beta_3\,,\nonumber\\
J &\equiv \beta_1+2\,\xi\,\beta_2+\xi^2\beta_3\,,\nonumber\\
\rho_{m,f} &\equiv \frac{1}{\xi^3}\,
\left(
\beta_1 +3\,\xi\,\beta_2+3\,\xi^2\,\beta_3
\right)
\,,\nonumber\\
\Gamma & \equiv 
\xi\,\beta_1+\xi^2\,\beta_2 + r\,\xi^2(\beta_2+\xi\,\beta_3)
\,,
\end{align}
where $r$ quantifies the alignment of the light-cone defined by the $\tilde{f}$--metric with respect to $g$: 
\begin{equation}
r \equiv \frac{\dot{f}\,\sqrt{1-D\,f^2}}{\xi}\,.
\end{equation}
With this definition, we can also express the time derivative of the ratio of the two scale factors as
\begin{equation}
\dot{\xi} = \xi\left[\frac{\sqrt{\kappa}\,r}{a\,\sqrt{1-\frac{a^2\xi^2D}{\kappa}}}-H\right]\,.
\end{equation}

We now calculate the background equations of motion by varying with respect to $N$ (which can be re-introduced by rescaling all time derivatives as well as the volume element), $a$, $f$ and matter perturbations, obtaining:
\begin{align}
3\,G\,\left[\left(H+\frac{\dot{G}}{2\,G}\right)^2-\frac{\kappa}{a^2}\right] &=
\frac{\rho_\chi}{\mpl^2}+m^2\,\rho_{m,g}\,,
\label{eq:EQN}
\\
-2\,G\,\left[\partial_t\left(H+\frac{\dot{G}}{2\,G}\right)+\frac{\kappa}{a^2}\right] + \dot{G}\,\left(H+\frac{\dot{G}}{2\,G}\right) &= \frac{\rho_\chi+p_\chi}{\mpl^2}+m^2 J\,\xi\,(1-r)\,,
\label{eq:EQA}
\\
\dot{\rho}_\chi &=-3\,H\,(\rho_\chi+p_\chi)\,,
\label{eq:EQC}
\end{align}
where $\rho_\chi$ and $P_\chi$ are the energy density and pressure of the matter fluid. We also defined $c_s$ as the propagation speed of the fluid. The equation of motion for the St\"uckelberg fields on the other hand is given by 
\begin{equation}
3\,m^2\,J\,\xi(1-r)\,\left(H+\frac{\dot{G}}{2\,G}\right)+m^2\,\dot{\rho}_{m,g}-\frac{\dot{G}}{2\,G}\left(\frac{\rho_\chi-3\,p_\chi}{\mpl^2}+4\,m^2\,\rho_{m,g}\right)=0\,.
\label{eq:EQS}
\end{equation}
In standard dRGT, the analogue of this equation forces the function $J$ to vanish around self-accelerating backgrounds. In our case, this is no longer true. In addition to the effect of the conformal factor $\dot{G}$, there is also the effect of the generalized mass terms encoded in $\dot\rho_{m,g}$. 
To make the latter effect explicit, we rewrite Eq.\eqref{eq:EQS} in terms of derivatives with respect to $\phi^a\phi_a$
\begin{equation}
3\,m^2\,J\,r\,\xi\left(\sqrt{1-\frac{a^2\xi^2D}{\kappa}}H-\frac{\sqrt{\kappa}}{a}\right)+\frac{2\,m^2 a\,r\,\xi^2}{\sqrt{\kappa}} \rho_{m,g}^{(1)}-\frac{a\,r\,\xi^2\,G'}{\sqrt{\kappa}\,\mpl^2\,G}\left[\rho_\chi -3\,p_\chi+m^2 \mpl^2\left(3\,J\,\xi\,(r-1)+4\,\rho_{m,g}\right)
\right]=0\,,
\label{eq:EQSprime}
\end{equation}
where $G' = G'(\phi^a\phi_a)$ and we defined
\begin{equation}
\rho_{m,g}^{(1)} \equiv \beta_0' +3\,X\,\beta_1' +3\,X^2\beta_2'+X^3\beta_3'\,.
\end{equation}
From Eq.~\eqref{eq:EQSprime}, we see that in standard dRGT, $D=0$, both $G$ and $\beta_n$ are constant, thus we either have $H = \sqrt{\kappa}/a$ (i.e. physical metric is Minkowski in open chart) or $J=0$ ~\cite{Gumrukcuoglu:2011ew}.

In the perturbation calculation, we solve the set of background equations \eqref{eq:EQN}--\eqref{eq:EQS} for $\rho_{m,g}$, $\dot{H}$, $\dot{\rho}$ and $\dot{\rho}_{m,g}$ and evaluate the quadratic action on-shell.

\subsubsection{Tensor sector}
For the tensor modes, one can obtain the action quadratic in perturbations as (after expanding in harmonics)
\begin{equation}
S^{(2)} = \frac{\mpl^2}{8}\,\int d^3k\, dt \,a^3\,G\,\left[
|\dot{\gamma}|^2-\left(\frac{k^2-2\,\kappa}{a^2} +\frac{m^2\Gamma}{G}\right)|\gamma|^2
\right]\,.
\end{equation}
The dispersion relation for the canonical mode is
\begin{align}
\omega_T^2 &= \frac{k^2-2\,\kappa}{a^2} +\frac{m^2\Gamma}{G} - \frac{3\,H\,\dot{G}}{2\,G}+\frac{\dot{G}^2}{4\,G^2}-\frac{\ddot{G}}{2\,G}\,.
\label{eq:tensordispersionFULL}
\end{align}
In contrast to the Horndeski theory in which the first derivative of the scalar field is non-minimally coupled to gravity \cite{Horndeski:1974wa, Kobayashi:2011nu}, the propagation speed of the tensor mode is the same as the speed of light, and the non-minimal coupling simply shift the mass of graviton.

\subsubsection{Vector sector}
For the vector modes, we first solve for the shift perturbations
\begin{equation}
B_i = \left[1+\frac{2\,a^2\xi}{G\,(k^2+2\,\kappa)}\,\frac{m^2J}{1+r}\right]^{-1}\,\frac{a\,\dot{E}_i}{2}\,.
\end{equation}
Substituting it back to the action, we find the following form for the reduced action
\begin{equation}
S^{(2)} = \frac{\mpl^2}{8}\,\int d^3k\, dt \,a^3\,\mathcal{T}_V\,\left[
|\dot{E_i}|^2-\left(c_V^2 \,\frac{k^2+2\,\kappa}{a^2} + \frac{m^2\,\Gamma }{G}\right)
|E_i|^2
\right]\,,
\end{equation}
where the kinetic term is
\begin{equation}
\mathcal{T}_V \equiv \left(
\frac{2}{G\,(k^2+2\,\kappa)}+\frac{r+1}{m^2 a^2\,\xi\,\,J}
\right)^{-1}\,,
\end{equation}
and the propagation speed is given by
\begin{equation}
c_V^2 = \frac{\Gamma(1+r)}{2\,\xi \,J}\,.
\end{equation}
In the UV, the kinetic term is
\begin{equation}
\mathcal{T}_V \Big\vert_{k\to\infty} = \frac{m^2 a^2  \xi J}{r+1}\,.
\end{equation}

In the dRGT limit, i.e. $G\to 1$\,,$J\to 0$, $D\to 0$, the sound speed reduces to
\begin{equation}
c_v^2 \Big\vert_{dRGT} = \frac{3\,m^2\,\mpl^2H^2\Gamma}{2\,\xi^2[G'(\rho_\chi-3\,P_\chi+4\,m^2\mpl^2\,\rho_{m,g})-2\,m^2\mpl^2\, \rho_{m,g}^{(1)}]} \to \infty
\end{equation}
where we assumed 
\begin{equation}
\mathcal{O}\left(\frac{G^{(n)}}{G} \right)\sim \mathcal{O}\left(\frac{\rho_{m,g}^{(n)}}{\rho_{m,g}}\right)\sim \mathcal{O}\left(\frac{\Gamma^{(n)}}{\Gamma}\right) \ll 1\,,~~~~{\rm for}~~n\ge 1\,,
\end{equation}
and $H \gg \sqrt{\kappa}/a$.

\subsubsection{Scalar sector}
\label{sec:scalar-conformal}
Finally, we calculate the quadratic action for scalar perturbations $(\Phi, \psi, B, E, \delta\chi)$. Two of these scalar perturbations are non-dynamical, and we can integrate out $\Phi$ and $B$. One of the remaining variables corresponds to the BD ghost. Then, one can check that the determinant of the remaining kinetic matrix composed of $\psi$, $E$, and $\delta\chi$ vanishes, and this implies the absence of the BD ghost. We then introduce the new variable $\tilde{\delta \chi}$ to remove the would-be BD ghost, 
\begin{equation}
\tilde{\delta \chi} \equiv \psi +\frac{k^2}{6}\,E -\frac{\mpl}{\dot{\chi}_0}\left(H + \frac{\dot{G}}{2\,G}\right)\,\delta\chi\,.
\label{redifinition:chi}
\end{equation}
Replacing $\delta\chi$ in favor of $\tilde{\delta \chi}$, the resulting action becomes independent of $\dot{\psi}$. We then integrate out $\psi$ and end up with an action containing only two degrees of freedom $\tilde{\delta \chi}$ and $E$. One of these corresponds to the matter perturbations while the other is the scalar polarization of graviton. Then, the quadratic action in the scalar sector can be formally written as 
\ba
S_S^{(2)} = {\mpl^2 \over 2} \int d^3 k\, dt\, a^3 \Big(
{\dot \Psi}^\dagger {\cal K} {\dot \Psi}
+ {1\over 2} {\dot \Psi}^\dagger {\cal G} {\Psi}
+ {1\over 2} { \Psi}^\dagger {\cal G}^T {\dot \Psi}
- { \Psi}^\dagger {\cal M} { \Psi}
\Big) \,,
\label{action:quadraticS}
\ea
where $\Psi \equiv (E, \tilde{\delta \chi})$ and ${\cal K}$, ${\cal G}$ and ${\cal M}$ are the real $2\times 2$ kinetic, mixing, and  mass matrices, and ${\cal K}$ and ${\cal M}$ are symmetric. 
The eigenvalues of the kinetic matrix in the subhorizon limit $k \to \infty$ yield the conditions to avoid ghost-like instability. These can be determined as 
\ba
\kappa_1= {\cal K}_{11} \,, \qquad \kappa_2 = {\det {\cal K} \over {\cal K}_{11}} \,.
\ea

The eigenvalues of the kinetic matrix in the subhorizon limit yield the conditions to avoid ghost-like instability. The first eigenvalue in this limit is simply
\begin{equation}
\kappa_1 = \frac{\rho_\chi+P_\chi}{c_\chi^2\left(H+\frac{\dot{G}}{2\,G}\right)^2}
+\mathcal{O}\left(\frac{k}{a\,H}\right)^{-2}\,,
\end{equation}
and it can be identified as the matter perturbation. It is not a ghost as long as the null energy condition is satisfied. The second eigenvalue is more complicated, but in subhorizon limit it can be written as
\begin{align}
\kappa_2 =& \frac{m^2r\,J^2\,\xi^2}{2\,G} + \frac{\kappa\,J\,r\,\xi}{a^2}+\left(H+\frac{\dot{G}}{2\,G}\right)^2(2\,\Gamma+J\,\xi)
+\frac{\rho_\chi+P_\chi}{2\,m^2\mpl^2}\left[
\frac{m^2\,J\,r\,\xi}{G}+(1-3\,c_s^2)\frac{\dot{G}}{G}\left(H+\frac{\dot{G}}{2\,G}\right)
\right]
\nonumber\\
&-\xi\,\left(H+\frac{\dot{G}}{2\,G}\right)\left(
4\,H\,J+\frac{J\,\dot{\xi}}{\xi}
+\dot{J}\right)+\mathcal{O}\left(\frac{k}{a\,H}\right)^{-2}
\label{eq:kinetic-conformal}
\end{align}
At this point, we see that the sub-horizon limit and dRGT limit do not commute. Sending $J,~\dot{J},~\dot{G}\to 0$
\begin{equation}
\kappa_2\Big\vert_{k\to\infty, ~dRGT} =2\,H^2\,\Gamma\,.
\end{equation}
If we reverse the order of the limits, we then get
\begin{equation}
\kappa_2\Big\vert_{dRGT,~k\to\infty} \to 0\,.
\end{equation}
This is compatible with what was found in the generalized dRGT without conformal transformation \cite{Kenna-Allison:2019tbu}.

Finally, we calculate the sound speeds of the scalar degrees of freedom. Unlike the minimally coupled case where one sound speed coincides with $c_s$, this is no longer true when non-minimal coupling is introduced. The full expression for the scalar sound speeds are reported in Appendix \ref{app:NMLsound}. 
As in the vector sector, in the dRGT limit, these reduce to
\begin{equation}
C_1^2 \Bigg\vert_{dRGT} = c_s^2\,,\qquad
C_2^2\Big\vert_{dRGT} = \frac{2\,m^2\,\mpl^2H^2\Gamma}{\xi^2\left[G'\left(\rho-3\,P+4\,m^2\mpl^2\,\rho_{m,g}\right)-2\,m^2\mpl^2\,\rho_{m,g}^{(1)}\right]}
=\frac{4}{3}c_v^2 \Big\vert_{dRGT} \to \infty
\,.
\end{equation}
As expected the strong coupling problem of the constant mass theory manifests itself as sound speed that diverge in the exact dRGT limit.

All of the expressions for this theory recovers the results of Ref.\cite{Kenna-Allison:2019tbu} in the limit $G\to 1$ and $\dot{\xi}\to \xi\left(\sqrt{\kappa}\,r/a-H\right)$ (or equivalently, $D\to 0$).

\subsection{Projected massive gravity theory}
In this subsection, we study the action \eqref{action:NMP} up to the quadratic mass terms, 
\ba
S
&=&
\int {\rm d}^4 x \sqrt{-g} \frac{\mpl^2}{2}\Biggl[ G \,R  + {6 G_X^2 \over G} [Y] 
+ m^2 \Big( A  [Z] +B_1  [Z]^2 + B_2 [Z^2] \Big)\Biggr] + S_{\rm m}  [g, \chi] \,,
\label{actionC:projected}
\ea
where $G$, $A$, $B_1$, and $B_2$ are functions of $X$.

\subsubsection{Background equations}
The background equations for gravity are given by
\ba
3G \left[\left(H+ {\Gdot \over 2G}\right)^2 - {\kappa \over a^2}\right]&=&
{\rho_\chi \over \mpl^2}+  {\rho_g\over \mpl^2} \,, \label{EOMn:NMP}\\
-2 G\left[
\p_t \left(H+ {\Gdot \over 2G}\right) + {\kappa \over a^2}
\right]
+ \Gdot \left( H+ {\Gdot \over 2G}\right) 
&=& {\rho_\chi + p_\chi \over \mpl^2}+ {\rho_g + p_g \over \mpl^2}
\label{EOMa:NMP}
\ea
where we introduced the effective density and pressure for the mass terms as
\ba
\rho_g &=& -{3\over 2} \mpl^2 m^2\xi^2 \left( A + (3 B_1 + B_2)\xi^2\right) \,,\\
p_g &=& {1\over 2} \mpl^2 m^2 \xi^2 \left( A - (3 B_1 + B_2)\xi^2\right) \,.
\ea
The equation of motion of the \Stuckelberg fields is given by 
\ba
\dot{\rho}_g + 3H (\rho_g +p_g)  - {\dot{G} \over2 G} \left(\rho_g-3p_g + \rho_\chi -3 p_\chi \right)  
=0 \,.
\label{EOMf:NMP}
\ea
Again, three of these equations are independent, and 
we solve these equations \eqref{EOMchi}, \eqref{EOMa:NMP}, and \eqref{EOMf:NMP} for $\dot{H}, \dot{ \rho_g}$, and $\rho_\chi$ for deriving the quadratic action.

\subsubsection{Tensor sector}
For the tensor modes, one can obtain the action quadratic in perturbations as 
\begin{equation}
	S^{(2)}_T = \frac{\mpl^2}{8}\,\int d^3k\, dt \,a^3\,G\,\left[
	|\dot{\gamma}|^2-\left(\frac{k^2-2\,\kappa}{a^2} 
	+\frac{M_{\rm GW}^2}{\,G}
		\right)|\gamma|^2
	\right]\,.
\end{equation}
The dispersion relation for the canonical mode is
\begin{align}
	\omega_T^2 &= \frac{k^2-2\,\kappa}{a^2}  
	+\frac{M_{\rm GW}^2}{G}
	- \frac{3\,H\,\dot{G}}{2\,G}+\frac{\dot{G}^2}{4\,G^2}-\frac{\ddot{G}}{2\,G}\,,
\end{align}
where we defined the mass of the tensor modes in the minimal coupling case, 
\ba
M_{\rm GW}^2 = \frac{2(\rho_g + p_g -2m^2 \mpl^2 B_2 \xi^4)}{\mpl^2} \,.
\ea
As in the theory \eqref{action:NMMG-re}, 
the propagation speed of the tensor mode is exactly the same as the speed of light.

\subsubsection{Vector sector}
Next, we expand the action up to quadratic order in the vector sector. Since the shift perturbation is non-dynamical,  we first solve the constraint equation as 
\begin{equation}
	B_i = 	\frac{\mpl^2 G(k^2 +2\kappa) }{2\mpl^2 G(k^2+2\kappa)+4a^2 (\rho_g+p_g)}
	a\,\dot{E}_i\,.
\end{equation}
Plugging this back into the action, we find the following form for the reduced action
\begin{equation}
	S^{(2)}_V = \frac{\mpl^2}{8}\,\int d^3k\, dt \,a^3\,\mathcal{T}_V\,\left[
	|\dot{E_i}|^2-\left(c_V^2 \,\frac{k^2+2\,\kappa}{a^2} 
	+ \frac{M_{\rm GW}^2}{G}
		\right)
	|E_i|^2
	\right]\,,
\end{equation}
where the kinetic coefficients is given by

\begin{equation}
	\mathcal{T}_V \equiv \frac{a^2 G(k^2 +2\kappa) (\rho_g+p_g)}{\mpl^2 G(k^2+2\kappa)+2a^2 (\rho_g+p_g)} \,,
\end{equation}
and the propagation speed for the vector mode is given by

\begin{equation}
	c_V^2 = \frac{\mpl^2 M_{\rm GW}^2}{2(\rho_g+p_g)} \,.
\end{equation}
The propagation speed of the vector polarization modes can in general differ from the speed of light. 
From this expression, we find that the ghost instability can be avoided when $\rho_g + p_g > 0$. 
In addition, the gradient instability in the vector sector is absent if $M_{\rm GW}^2 \geq 0$, and it coincides with the condition for avoiding the tachyonic instability in the tensor modes, when $\dot{G}=0$.

\subsubsection{Scalar sector}
Now, let us move on to the scalar perturbations. As in the case of the previous theory class \eqref{action:NMMG-re}, we can integrate out the non-dynamical variables $\Phi$, $B$, and $\psi$ by using the same $\tilde{\delta\chi}$ defined by \eqref{redifinition:chi}, and the reduced action can be written in the form of \eqref{action:quadraticS}.
The eigenvalues are given by
\begin{align}
	\kappa_1 =& \frac{1}{\left(H+\frac{\dot{G}}{2\,G}\right)^2}
	\left[
	\frac{2\,c_\chi^2}{\rho_\chi+P_\chi}-\frac{(1-3\,c_\chi^2)^2\,\dot{G}^2}{G^2\left(H+\frac{\dot{G}}{2\,G}\right)}
	\left(\frac{\dot{G}}{G}\left[ 9\,c_\chi^2\,\rho_\chi-3(4-3\,c_\chi^2)P_\chi+\rho_g-3\,P_g\right]+2\,\left[
	H\,(\rho_g+9\,P_g)+3\,\dot{P}_g
	\right]\right)^{-1}
	\right]^{-1}
	\nonumber\\
	&+\mathcal{O}\left(\frac{k}{a\,H}\right)^{-2}\,,
	\nonumber\\
	\kappa_2 = & \frac{a^2(\rho_g+P_g)}{8}k^2 - \frac{3\,a^4\,(\rho_g+P_g)^2}{16\,\mpl^2G} + \mathcal{O}\left(\frac{k}{a\,H}\right)^{-2}\,.
\end{align}
The first eigenvalue $\kappa_1$ can be identified as the matter perturbation which can be easily seen when $G=1$, 
\begin{equation}
	\kappa_1 = \frac{\rho_\chi+p_\chi}{2c_\chi^2 H ^2 }
	+\mathcal{O}\left(\frac{k}{a\,H}\right)^{-2}\,.
\end{equation}
Therefore, the scalar graviton is free from ghost when $\rho_g + p_g >0$.
The full expression of the propagation speeds are summarized in the Appendix \ref{app:NMLsound}. 

\subsubsection{Concrete model}
In this subsection, we give a concrete model in the projected theory \eqref{action:NMP}. Let us first choose the simplest functions:
\ba
G=1,\qquad A=a_1, \qquad B_1 =  b_1, \qquad B_2 = b_2 \,,
\ea
where $a_1, b_1$, and $b_2$ are constants.
Then, the equation of motion for $f$ gives the constraint equation, 
\ba
\xi (H \xi + \dot{\xi}) \left[a_1+2(3b_1+b_2)\xi^2\right]=0 \,.
\ea 
Assuming that $\xi$ is nonzero, 
the first solution $H \xi + \dot{\xi}=0$ gives $\xi \propto 1/a$. In this case, $\rho_g$ behaves as the sum of the spacial curvature and the radiation, so we do not discuss this solution here. The second one gives 
\ba
\xi = \pm \sqrt{\frac{-a_1}{2(3b_1 + b_2)}} \,.
\label{xi0}
\ea
In order for $\xi$ to be real, we impose $a_1/(3b_1 + b_2) <0$. 
Since $\xi$ is a constant, $\rho_g$ and $p_g$ now becomes constants, 
\ba
\rho_g = -p_g = \frac{3a_1^2 m^2 \mpl^2}{8(3b_1 + b_2)} \,.
\ea
Since the mass term exactly behaves as a cosmological constant, the kinetic terms of the vector and scalar graviton modes vanish in this case ($\rho_g+p_g=0$). This strong coupling behavior can be avoided once the $X$-dependence in the arbitrary functions $G$, $A_1$, $B_1$, or $B_2$ is taken into account. Here, let us consider the  non-minimal coupling case,  
\ba
G = 1 + g \, m^2X  \,,
\ea
where we assume $g \ll 1$.  Then, we expand $\xi$ in terms of this small parameter $g$, 
\ba
\xi = \xi_0 + g \xi_1 + {\cal O}(g^2) \,.
\ea
Here, $\xi_0$ is given by the positive sign of \eqref{xi0},
\ba
\xi_0 = \sqrt{\frac{-a_1}{2(3b_1 + b_2)}} \,. 
\ea
From the equation of motion for $f$, we obtain
\ba
\xi_1 =- \frac{a^2 \xi_0}{6 a_1 \kappa  \mpl^2} (\rho_\chi- 3p_\chi -3 a_1 m^2 \mpl^2\xi_0^2) \,,
\ea
and the energy density and pressure can be then expressed as
\ba
\rho_g &=& -{3\over 4} a_1 m^2 \mpl^2 \xi_0^2 + {\cal O}(g^2) \,,\\
p_g &=& {3\over 4} a_1 m^2 \mpl^2\xi_0^2 + 2 a_1 m^2 \mpl^2 \xi_0 \xi_1 g
+ {\cal O}(g^2)  \,.
\ea
Then, the background equations become
\ba
3\left[\left(1-3 g  m^2 \xi_0^2\frac{a^2 }{\kappa}\right)H^2 -{\kappa \over a^2}\right]
&\simeq& {\rho_\chi \over \mpl^2} +{ {\tilde \rho}_g  \over \mpl^2} \,,\\
-2 \left[ \left(1-2g m^2  \xi_0^2 \frac{a^2 }{\kappa }\right) \dot{H} +\frac{\kappa}{a^2}
- g m^2 \xi_0^2  \frac{a^2 }{\kappa} H^2
\right]
&\simeq& {\rho_\chi + p_\chi \over \mpl^2} + \frac{{\tilde \rho}_g + {\tilde p}_g}{\mpl^2} \,,
\ea
where 
\ba
{\tilde \rho}_g &=& \rho_g -3 g \,m^2 \mpl^2 \xi_0^2 \,, \\
{\tilde p}_g &=& p_g + g \, m^2 \mpl^2\xi_0^2  \,.
\ea
Now we would like to derive the conditions for avoiding ghost and gradient instabilities. As mentioned the above, the positivity of the cosmological constant requires that $\rho_g>0$, and $\xi_0$ has to be real,  
\ba
a_1 < 0, \qquad 3b_1 + b_2 >0 \,.
\ea
All modes are ghost-free when $\rho_g + p_g >0$, which gives
\ba
\xi_1 g <0 \,.
\ea
Substituting the background solution, the propagation speed of the vector mode is given by  
\ba
c_V^2=
- \frac{ b_2 \xi_0^3}{a_1 \xi_1 g} + {\cal O} (g^0) \,,
\ea
and the propagation speeds of the scalar modes can be now simplified as
\ba
c_1^2 = c_\chi^2\,,\qquad c_2^2 = - \frac{4 b_2 \xi_0^3}{3a_1 \xi_1 g} + {\cal O} (g^0) \,.
\ea
Combining these conditions, we obtain
\ba
a_1 <0, \qquad 3b_1+b_2 > 0, \qquad 
\xi_1 g < 0, \qquad b_2 < 0 \,.
\label{GFCondition:projection}
\ea
Therefore, all modes are free of ghost and gradient instabilities when the conditions \eqref{GFCondition:projection} are satisfied.

\section{Summary}
\label{sec:summary}

In the present paper, we studied  a generalization of massive gravity with the broken translation invariance. Introducing a deformation to the fiducial metric ${\tilde f}_{\mu\nu} = (\eta_{ab} + D\,\phi_a\phi_b)\,\partial_\mu \phi^a\,\partial_\nu \phi^b$ is essential to find extended theories beyond the dRGT massive gravity. Starting with arbitrary mass functions, we found two potential ways to avoid the BD ghost. The first case is the extension of the generalized massive gravity and any detuning of the quadratic dRGT potential requires a non-minimal coupling with curvature. The action for this theory is given by
\ba
S = \int d^4x\,\sqrt{-{g}}\frac{\mpl^2}{2} \left[ 
GR+ 
{6 G_X^2 \over G} [Y]
-2m^2\sum_{n=0}^3\,{\beta}_n(X) \,e_n\left(\sqrt{{g}^{-1}\tilde{f}}\right)\right] + S_{\rm m}  [g, \psi] \,. 
\ea

The second theory can be constructed using the fiducial metric ${\bar f}_{\mu\nu} = P_{ab}\,\partial_\mu \phi^a\,\partial_\nu \phi^b$, where we use the projection tensor $P_{ab} = \eta_{ab} - {\phi_a\phi_b /X}$, which manifestly eliminates one of the \Stuckelberg fields along $\phi^a$. 
The action for the projected theory is given by
\ba
S=
\int {\rm d}^4 x \sqrt{-g} \frac{\mpl^2}{2}\Biggl[ G \,R  + {6 G_X^2 \over G} [Y] 
+ m^2  \,{\cal U} \bigl(X, [Z], [Z^2], [Z^3] \bigr)
\Biggr]+ S_{\rm m}  [g, \psi]\,,
\label{eq:projected-conclusions}
\ea
where $Y$ and $Z$ are defined in \eqref{eq:defWY}. 
In the form that we proposed, this theory can also have the same non-minimal coupling, but the mass term is no longer of the form of the dRGT potential terms. In addition, the potential term is an arbitrary function of $X$ and $[Z^n]$. We have systematically proved the absence of the BD ghost in this theory. The projected theory action \eqref{eq:projected-conclusions} is actually not the most general, since there remains some freedom to include further non-minimal coupling without generating the BD ghost. 
For instance, the term $G^{\mu\nu}\bar{f}_{\mu\nu}$ was considered in Ref.\cite{Lin:2013aha}. The possibility of other non-minimal coupling terms can be easily seen by considering general disformal transformations of the metric tensor $g_{\mu\nu}\to \tilde{g}_{\mu\nu} =C g_{\mu\nu} +D\, Z_{\mu\nu}+E\, (Z^2)_{\mu\nu}+F\,(Z^3)_{\mu\nu}$, where all coefficients are functions of $X$, $[Z]$, $[Z^2]$, $[Z^3]$. 
Such a transformation would generate the $G^{\mu\nu}\bar{f}_{\mu\nu}$ coupling, as well as many others, and we will report this in a later study. 

We have then studied open-FLRW cosmologies of these obtained theories. In both cases, 
all perturbations are free of ghost and gradient instabilities. In addition, we have found that the structure of the non-minimal coupling does not change the propagation speed of the tensor modes while the vector and scalar graviton propagates either subluminal or superluminal speed.

These new theories are the first ones where the kinetic term of a massive graviton is no longer Einstein-Hilbert term due to the non-minimal coupling, and it might bring a new phenomenology of large scale structure. For instance, the translation breaking will be manifested as time variation in coupling constants. For solar system tests, we expect that the theory \eqref{action: NMMG} exhibits Vainshtein mechanism due to its connection to generalized galileon theories in the decoupling limit. The phenomenology of this theory will be investigated in a future publication \cite{Kenna-Allison:2020}. Conversely, the theory \eqref{action:NMP} is disconnected from the dRGT construction, thus the existence and/or necessity of a screening mechanism needs to be confirmed.
%These new theories are the first ones where the kinetic term of a massive graviton is no longer Einstein-Hilbert term due to the non-minimal coupling, and 
%it might bring a new phenomenology of large scale structure.
%It would be also interesting to investigate if the Vainshtein mechanism is necessary in the projected theory, and if so, whether it can be realized.

\acknowledgments
Some of the calculations for cosmological perturbations have been performed using the xPand package \cite{Pitrou:2013hga}.
This work was supported in part by JSPS Grant-in-Aid for Scientific Research Nos. JP17K14276
 (RK). The work of AEG and KK has received funding from the European Research Council
(ERC) under the European Union's Horizon 2020 research and innovation programme (grant agreement No. 646702 ``CosTesGrav''). KK is supported by the UK STFC ST/S000550/1.

\appendix

\section{Degeneracy conditions around fixed backrounds}
\label{app:fixedbg-degeneracy}

Due to the complexity of the action \eqref{action:quadratic}, we here simplify the derivation of degeneracy conditions, instead of using $3+1$ decomposition.
An obvious first choice is the homogeneous and isotropic background. However, even for constant mass parameters, one cannot even deduce the full degeneracy conditions that yield the dRGT form. In this Appendix, we instead consider two backgrounds that have fewer symmetries than the cosmological background. We first consider a homogeneous background with broken isotropy, which yields explicit necessary conditions. We next study an inhomogeneous background, which gives tighter degeneracy conditions that we could not write down explicitly. However, we are able to check some options for the relations between functions. We summarize our findings at the end of the appendix, which form the basis of the conditions \eqref{eq:condition-transformation-functions}-\eqref{eq:condition-mass-functions} quoted in the main text.

The action we consider in this appendix is
\begin{equation}
	S = \int d^4x \sqrt{-g} \frac{\mpl^2 }{2} \Bigl[G(X)R[g]+F(X)[Y]
	+  A (X) [W]
	-2\,m^2 {\cal L}_{\rm mass}
	\Bigr]\,,
	\label{action:quadratic-appendix}\,
\end{equation}
where we defined $W$ and $Y$ in Eq.\eqref{eq:defWY}, while ${\cal L}_{\rm mass}$ is given in Eq.\eqref{eq:Lmass-deformed}.

\subsection{Degeneracy around anisotropic background}
\label{subsec:BianchiV}

We first start with the Bianchi type-V spacetime, which is the simplest anisotropic background that is compatible with uniform $\phi^a\phi_a$. The physical metric is given by
\begin{equation}
	ds^2 = - dt^2 +a(t)^2 dx^2 + {\rm e}^{2\,\alpha\,x}\left(b(t)^2 dy^2+c(t)^2 dz^2\right)\,.
\end{equation}
The scalar field configuration is chosen to be
\begin{align}
	\phi^0 =& f(t)\left[\cosh(\alpha\,x)+\frac{\alpha^2(y^2+z^2)\,{\rm e}^{\alpha\,x}}{2}\right]\,,\nonumber\\
	\phi^1 =& f(t)\left[\sinh(\alpha\,x)-\frac{\alpha^2(y^2+z^2)\,{\rm e}^{\alpha\,x}}{2}\right]\,,\nonumber\\
	\phi^A =& f(t)\,\alpha\,x^A\,{\rm e}^{\alpha\,x}\,,
\end{align}
where $A=2,3$. With this choice, we have $X=\eta_{ab}\phi^a\phi^b = -f^2$ and
\begin{equation}
	f_{\mu\nu}dx^\mu dx^\nu = -\dot{f}^2(t)dt^2 + \alpha^2 f(t)^2dx^2+\alpha^2 f(t)^2 {\rm e}^{2\,\alpha\,x}\left(dy^2+dz^2\right)\,.
\end{equation}
This is simply the Minkowski metric written in a chart that is compatible with the Bianchi type-V form.
We can also define the fiducial metric obtained from a transformed field space metric \eqref{eq:disformalreference}
\begin{equation}
	\tilde{f}_{\mu\nu, I}dx^\mu dx^\nu = -(C_I-f^2\,D_I)\dot{f}^2dt^2 + \alpha^2 C_I\,f(t)^2dx^2+\alpha^2 f(t)^2 C_I\,{\rm e}^{2\,\alpha\,x}\left(dy^2+dz^2\right)\,.
\end{equation}
Then, we have the diagonal matrix
\begin{equation}
	(g^{-1}{\tilde f}_I)^\mu_{\;\;\nu} = 
	\left(
	\begin{array}{cccc}
		(C_I+D_IX)\dot{f}^2 & 0 & 0 & 0 \vspace{1mm}\\
		0 & \displaystyle{\alpha^2 C_I\,\,\frac{f^2}{a^2}} & 0 & 0\vspace{1mm}\\
		0 & 0 & \displaystyle{\alpha^2 C_I\,\,\frac{f^2}{b^2}} & 0 \vspace{1mm}\\
		0 & 0 & 0 & \displaystyle{\alpha^2 C_I\,\,\frac{f^2}{c^2}}
	\end{array}
	\right)\,.
\end{equation}
We can then evaluate the action \eqref{action:quadratic} for this background. We vary the action in the minisuperspace approximation $\mathcal{S}$ with respect to the variables  $a$, $b$, $c$ and $f$, and obtain four dynamical equations 
\begin{equation}
	E_n \equiv (E_a, E_b, E_c ,E_f)\,,\qquad
	E_a \equiv \frac{\delta \mathcal{S}}{\delta a}\,,\qquad
	E_b \equiv \frac{\delta \mathcal{S}}{\delta b}\,,\qquad
	E_c \equiv \frac{\delta \mathcal{S}}{\delta c}\,,\qquad
	E_f \equiv \frac{\delta \mathcal{S}}{\delta f}\,.
\end{equation}
Since these equations of motion contain the second time derivative of $f(t)$, to ensure the absence of the BD ghost, $f(t)$ should be non-dynamical. Therefore, we require that the kinetic matrix
\begin{equation}
	K_{mn}  = \frac{\partial E_m}{\partial \ddot{q}_n}\,,
\end{equation}
is degenerate. Here we have defined the variables as $q_n \equiv (a,b,c,f)$.
Assuming $G\neq 0$, the determinant of the kinetic matrix is given by 
\begin{align}
	\det K =&
	\mathcal{D}_1 +6\,\dot{f}^2 \mathcal{D}_2-2\,m^2\,\alpha\,f\left(\frac{1}{a}+\frac{1}{b}+\frac{1}{c}\right)\left(3\,\dot{f}\,\mathcal{D}_3+\mathcal{D}_5\right) -6\,m^2 \dot{f}\,\mathcal{D}_4
	\nonumber\\
	& -4\,m^2\alpha^2\,f^2\,\left[\left(\frac{1}{a\,b}+\frac{1}{a\,c}+\frac{1}{b\,c}\right)\mathcal{D}_6 + \left(\frac{1}{a^2}+\frac{1}{b^2}+\frac{1}{c^2}\right)\,\mathcal{D}_7\right]\,,
\end{align}
where
\begin{align}
	\mathcal{D}_1=&
	A + F\,X-6\,X\,\frac{G_X^2}{G}-2\,m^2\Big[(C_{\gamma_1}+X\,D_{\gamma_1})\gamma_1+(C_{\gamma_2}+X\,D_{\gamma_2})\gamma_2\Big]\,,\nonumber\\
	\mathcal{D}_2=&
	X^2\,B_1 -2\,m^2 \Big[
	(C_{\sigma_1}+XD_{\sigma_1})^2\sigma_1
	+(C_{\sigma_2}+XD_{\sigma_2})^2\sigma_2
	+(C_{\sigma_3}+XD_{\sigma_3})^2\sigma_3
	+(C_{\sigma_4}+XD_{\sigma_4})^2\sigma_4
	+(C_{\sigma_5}+XD_{\sigma_5})^2\sigma_5
	\Big]
	\,,\nonumber\\
	\mathcal{D}_3=&
	4\,\sqrt{C_{\sigma_1}}(C_{\sigma_1}+XD_{\sigma_1})^{3/2}\sigma_1
	+2\,\sqrt{C_{\sigma_2}}(C_{\sigma_2}+XD_{\sigma_2})^{3/2}\sigma_2
	+\sqrt{C_{\sigma_4}}(C_{\sigma_4}+XD_{\sigma_4})^{3/2}\sigma_4
	\,,\nonumber\\
	\mathcal{D}_4=&
	(C_{\delta_1}+XD_{\delta_1})^{3/2}\delta_1
	+(C_{\delta_2}+XD_{\delta_2})^{3/2}\delta_2
	+(C_{\delta_3}+XD_{\delta_3})^{3/2}\delta_3
	\,,\nonumber\\
	\mathcal{D}_5=&
	3\,\sqrt{C_{\delta_1}}(C_{\delta_1}+XD_{\delta_1})\delta_1
	+\sqrt{C_{\delta_2}}\,(C_{\delta_2}+XD_{\delta_2})\delta_2
	\,,\nonumber\\
	\mathcal{D}_6=&
	6\,C_{\sigma_1}(C_{\sigma_1}+XD_{\sigma_1})\sigma_1
	+C_{\sigma_2}(C_{\sigma_2}+XD_{\sigma_2})\sigma_2
	\,,\nonumber\\
	\mathcal{D}_7=&
	3\,C_{\sigma_1}(C_{\sigma_1}+XD_{\sigma_1})\sigma_1
	+C_{\sigma_2}(C_{\sigma_2}+XD_{\sigma_2})\sigma_2
	+C_{\sigma_3}(C_{\sigma_3}+XD_{\sigma_3})\sigma_3
	\,.
	\label{eq:bianchi5conditions}
\end{align}
For the kinetic matrix to be non-invertible, all seven of these functions should be zero,
\ba
\mathcal{D}_1=\mathcal{D}_2=\mathcal{D}_3=\mathcal{D}_4=\mathcal{D}_5=\mathcal{D}_6=\mathcal{D}_7=0 \,.
\label{DC:BianchiV}
\ea
Here, the tadpole term, $\beta(X) \,[Q_{\beta}]$, in \eqref{eq:Lmass-deformed} does not contribute to the kinetic matrix since it is linear in $\dot f$ in this background.
One can confirm that when the above conditions are imposed, the study of linear perturbations does not reveal any new information on degeneracy.
Note that we can obtain the dRGT tuning in the translation invariant case, $F=A=D_I=0$ and $G = C_I =1$.
In a FLRW background, where $\alpha=0$ and $a=b=c$, the conditions $\mathcal{D}_6=0$ and $\mathcal{D}_7=0$ are combined into a single condition, that is, $\mathcal{D}_6+\mathcal{D}_7=0$. This demonstrates that the FLRW background is not adequate to reveal all of the dRGT tuning. 
Although these conditions are sufficient to eliminate the BD ghost in the Bianchi type-V background, the BD ghost reappears in more general backgrounds as we show in the next subsection.

\subsection{Degeneracy around inhomogeneous background}
So far, we have considered degeneracy conditions around a homogeneous but anisotropic background. It is therefore a legitimate question whether these conditions are sufficient to ensure nonlinear degeneracy (or equivalently, degeneracy around arbitrary backgrounds). 
We here consider a fixed physical metric given by \cite{deRham:2011rn,Kugo:2014hja}
\begin{equation}
	g_{\mu\nu} dx^\mu dx^\nu= -dt^2 +h_{ij} (dx^i + N^i dt)(dx^j + N^j dt)\,,
\end{equation}
where we use spherical coordinates $dx^i = (dr,\,d\theta,\,d\phi)$ and consider flat hypersurfaces $h_{ij} = {\rm diag}(1,\, r^2,\, r^2\,\sin^2\theta)$. We also use a shift vector that is aligned with the radial direction $N^i = (l, 0, 0)$. As for the scalar field configuration, we consider
\begin{align}
	\phi^0 &=f(t)\,\sqrt{1+\kappa r^2} + \delta\phi^0\,,\nonumber\\
	\phi^1 &=f(t)\,\sqrt{\kappa }r\,\sin\theta\,\cos\phi + \delta\phi^1\,,\nonumber\\
	\phi^2 &=f(t)\,\sqrt{\kappa }r\,\sin\theta\,\sin\phi + \delta\phi^2\,,\nonumber\\
	\phi^3 &=f(t)\,\sqrt{\kappa }r\,\cos\theta+ \delta\phi^3\,.
\end{align}
where $\delta \phi^a$ are perturbations. With this choice, the background $\tilde{f}_{\mu\nu}$ is diagonal
\begin{equation}
	{\tilde f}_{\mu\nu, I} = {\rm diag}\Big[-(C_I-f^2D_I)\dot{f}^2\,,\; \;
	\frac{\kappa \,C_I\,f^2}{1+\kappa \,r^2}\,,\;\;
	\kappa \,C_I\,r^2f^2\,,\;\;
	\kappa \,C_I\,r^2f^2\sin^2\theta \,\Big]
	+ {\cal O} (\delta \phi^a) \,,
\end{equation}
while the physical metric has a $2\times2$ non-diagonal block in the $(t,r)$ space. This example, although not necessarily corresponding to any solution of the equations of motion, nevertheless provides a background which is minimally nondiagonal, potentially revealing new degeneracy conditions not covered by \eqref{eq:bianchi5conditions}. Since the background is not a consistent parameterisation of the degrees of freedom, we need to look at the action quadratic in perturbations to deduce the conditions on degeneracy. 

At the background level, we have
\begin{equation}
	(g^{-1}\tilde{f}_I)^\mu_{\;\;\nu} = 
	\left(
	\begin{array}{cccc}
		(C_I-f^2D_I)\dot{f}^2 & \displaystyle{\frac{\kappa \,l\,C_I\,f^2}{1+\kappa \,r^2}} & 0 & 0 \vspace{1mm}\\
		-l\,(C_I-f^2D_I)\,\dot{f}^2 & \displaystyle{\frac{\kappa \,(1-l^2)\,C_I\,f^2}{1+\kappa \,r^2}} & 0 & 0\vspace{1mm}\\
		0 & 0 & \kappa \,C_I\,f^2 & 0 \vspace{1mm}\\
		0 & 0 & 0 & \kappa \,C_I\,f^2 
	\end{array}
	\right)
	+ {\cal O} (\delta \phi^a) \,.
\end{equation}
We start by diagonalising this tensor.
The background eigenvalues for the $(t,r)$ non-diagonal block are
\begin{equation}
	\lambda_{1,2}^I \equiv \frac{1}{2}\,\left[\frac{\kappa (1-l^2)C_I\,f^2}{1+\kappa \,r^2}\,+(C_I-f^2D_I)\,\dot{f}^2\right]\,\left(1\pm \sqrt{1-\frac{4\,\kappa (1+\kappa r^2)\,C_I\,f^2(C_I-f^2D_I)\,\dot{f}^2}{[\kappa (1-l^2)\,C_I\,f^2+(1+\kappa \,r^2)\,(C_I-f^2D_I)\,\dot{f}^2]^2}}\right)\,.
\end{equation}
We can find perturbation corrections to the eigenvalues of $g^{-1}\tilde{f}$ by solving
\begin{equation}
	\det (g^{-1}\tilde{f}_I - \mathbb{1}\,\ell)  = 0\,,
\end{equation}
perturbatively up to second order in perturbations. Unfortunately, this process is rather bulky for presentation. To simplify the process, we fix the angles $\theta=\pi/2$, $\phi=0$ and assume all perturbations are time dependent only, since we are eventually interested in terms quadratic in time derivatives. In the end, we formally have 
\begin{equation}
	(R^T g^{-1}\tilde{f} R)^\mu_{\;\;\nu} = \ell(\mu)\,\delta_{\nu\mu}\,,
\end{equation}
where $R$ is an orthogonal matrix and $\ell(\mu)$ denotes the $\mu$th eigenvalue. Then, observing that
\begin{equation}
	R^T\,\sqrt{g^{-1}\tilde{f}}\,R = \sqrt{R^Tg^{-1}\tilde{f} R}\,,
\end{equation}
we deduce that $\sqrt{\ell(\mu)}$ are the eigenvalues of $\sqrt{g^{-1}\tilde{f}}$, so they can directly be used when calculating the various traces.

Using Mathematica, we calculate the kinetic matrix and degeneracy conditions generated by only the mass terms, and hereafter we thus set $\dot{G}=F=A=0$. Note that the inclusion of the non-minimal coupling is justified by $3+1$ decomposition in section \ref{sec:3+1}. 
The expression of the degeneracy conditions is cumbersome, and it is difficult to solve these equations exactly in general. Thus, we here systematically assign random values to the function, and then we can confirm the degeneracy or non-degeneracy of the system to obtain the conditions\footnote{If there is degeneracy, it is not possible to conclusively demonstrate this using perturbative methods around fixed backgrounds.}.
Using this approach, we find that the following conditions complement \eqref{DC:BianchiV} :
\begin{align}
 C_{\gamma_1}=C_{\gamma_2}\,, \qquad
C_{\delta_1}=C_{\delta_2}=C_{\delta_3}\,,\qquad 
C_{\sigma_1}=C_{\sigma_2}=C_{\sigma_3}=C_{\sigma_4}=C_{\sigma_5}\,,
\end{align}
and
\begin{equation}
\frac{D_I}{C_I} = D(X)\qquad{\rm for ~ any~label~}I\,,
\end{equation}
where $D$ is a single function. These conditions imply that only a single field space metric is allowed, whilst the conformal factors $C_I$ can be absorbed in the definitions of the mass function. This can be seen as generalizing the original field space metric $\eta_{ab}$ to $\tilde{\eta}_{ab}(\phi^a)$.

\section{Derivation of the general action for the projected theory }
\label{app:NMP}
In this Appendix, we derive the action \eqref{action:NMP} in a systematic way. In contrast to the square root structure of the dRGT mass terms, we here construct mass terms by using $W$ and $Y$, defined in Eq.\eqref{eq:defWY}, as building blocks. 
Let us consider the most general mass terms up to quadratic order in $W$ and $Y$, 
\ba
{\cal L}_{\rm mass} &=&   B_1  [W]^2 + B_2 [W^2]+ B_3 [Y]^2+ B_4 [Y^2]+ B_5 [W][Y]+B_6 [W Y] \,,
\ea
where $B_i$ are function of $X $.
Then we consider the following action including the non-minimal coupling,  
\ba
S
&=&
\int {\rm d}^4 x \sqrt{-g} \frac{\mpl^2}{2}\Biggl[ G(X) R[g]+ F(X) [Y] + A(X) [W]+ m^2
{\cal L}_{\rm mass} \Biggr]\,.
\ea
After $3+1$ decomposition, the canonical momenta are given by 
\ba
\pi^a &=& {\delta {\cal L} \over \delta \phid_a} \notag \\
&=&
-4G_X K \phi^a -2 A \phid^a -2 F \phi^a \phi^b \phid_b 
+ 4(B_1 + B_2) \phid^a \phid_b \phid^b 
+2(B_5+B_6) (\phi^b \phi^c \phid^a \phid_b \phid_c + \phi^a \phi^b \phid_b\phid_c \phid^c)\notag \\
&&
+ 4(B_3 +B_4) \Bigl(\phi^a \phi^b \phi^c \phi^d \phid_b \phid_c \phid_d 
- \phi^a \phi^b \phi^c \phi^d \phid_b (D_\mu \phi_d) (D^\mu \phi_c)\Bigr)\notag \\
&&
-4B_1  \phid^a (D_\mu \phi_b) (D^\mu \phi^b)
-4B_2 \phid^b (D_\mu \phi_b) (D^\mu \phi^a)
-2B_5 \Bigl(\phi^a\phi^b \phid_b (D_\mu \phi_c) (D^\mu \phi^c) + \phi^b\phi^c \phid^a (D_\mu \phi_c) (D^\mu \phi_b)\Bigr)\notag \\
&&
-2B_6 \Bigl( \phi^b\phi^c \phid_b (D_\mu \phi_c) (D^\mu \phi^a)+ \phi^a\phi^b \phid^c (D_\mu \phi_c) (D^\mu \phi_b) \Bigr) \,,
\\
\pi^{\mu\nu} &=& {\delta {\cal L} \over \delta K_{\mu\nu}}
= 2G (K^{\mu\nu}-\gamma^{\mu\nu} K) -4G_X \gamma^{\mu\nu} \phi^a \phid_a  \,.
\ea
As in Sec.~\ref{sec:3+1}, we consider the linear combination of the canonical momenta 
\ba
\Psi \equiv \alpha_1 \phi^a \pi_a + \alpha_2 \gamma^{\mu\nu} \pi_{\mu\nu} \,,
\ea
where $\alpha_1$ and $\alpha_2$ are constants.
To ensure the existence of a primary constraint on arbitrary backgrounds, $\Psi$ should be independent of $K$, $\phi_a\phid^a$, $w_1 \equiv \phi^a\phid^b D_\mu \phi_a D^\mu \phi_b$, $w_2 \equiv \phi_a\phid^a D_\mu \phi_b D^\mu \phi^b$, and $w_3 \equiv \phi_a\phi_b \phi_c \phid^a D_\mu \phi^b D^\mu \phi^c$, i.e., 
\ba
\frac{\partial \Psi}{\partial (\phi_a\phid^a)} = 0\,, \qquad 
\frac{\partial \Psi}{\partial K} = 0 \,. \qquad
\frac{\partial \Psi}{\partial w_1} = 0\,, \qquad 
\frac{\partial \Psi}{\partial w_2} = 0\,, \qquad 
\frac{\partial \Psi}{\partial w_3} = 0\,, 
\ea
which gives the five conditions,
\ba
&& G_X X \alpha_1 + G \alpha_2 =0 \,, \qquad 
(A  + F X )\alpha_1 + 6 G_X \alpha_2 =0\,, \notag \\
&& B_5 + B_6 + 2(B_3 + B_4)X  =0 \,,\qquad
6B_1 + 4B_2 + 5B_5 X + 4 X (B_6 + (B_3 + B_4)X) =0\,,\notag \\
&&10 B_1 + 12 B_2 + X (7 B_5+ 8 B_6 + 4(B_3 + B_4 )X)  =0\,.
\ea
Solving these equations, we find 
\ba
F = {6 G_X^2 \over G} -{A \over X}, \quad 
B_6 =- {2B_2 \over X},\quad 
B_5 = -{2B_1 \over X}, \quad 
B_4 = - {B_1 +B_2 -B_3 X^2 \over X^2} \,.
\label{DCB}
\ea
As one can see, the degeneracy conditions for $G, A, F$ and $B_i$ do not mix, and this implies that they can be imposed at each order. 
In the translation invariant case ($B_{3,4,5,6}=0$ and $B_{1,2}=$const.), we have
$B_1=B_2=0$. Thus these translation-breaking terms are crucial to ensure the degeneracy. 
Once we impose these degeneracy conditions, the mass term ${\cal L}_{\rm mass}$ is characterized by only the arbitrary functions $B_1$ and $B_2$, while the $B_3$ term is canceled due to the above conditions. 
We can then rewrite the mass term in terms of only the projection tensor, and it can be written as ${\cal L}_{g} =  B_1  [Z]^2 + B_2 [Z^2]$. 
Thus, besides the non-minimal coupling and its counter terms, the mass terms can be described by the traces of the matrix $Z^\mu_{~\nu}$. 
With the same procedure, one can easily show that the cubic mass terms are described by all the possible combinations of the traces of $Z^\mu_{~\nu}$ with three arbitrary functions. The higher order extension can be also possible, and we finally arrive at the action \eqref{action:NMP}.

\section{Scalar sound speeds in the extended theories with non-minimal coupling}
\label{app:NMLsound}
In this section, we present the full expressions of the sounds speeds of scalar perturbations in the non-minimally coupled theories defined by the actions \eqref{action:NMMG-re} and \eqref{actionC:projected}. After the non-dynamical degrees of freedom are integrated out, the reduced scalar action contains two propagating degrees, one corresponding to the matter perturbations and the other to the scalar graviton polarization. Considering a monochromatic wave and taking the sub-horizon limit, we can solve the equations of motion, which can formally be reduced to the following algebraic equation for the sound speed $C$:
\begin{equation}
\left(\frac{C^2}{c_s^2}-1\right)(\mathcal{A}\,C^2-\mathcal{B})-\mathcal{D}=0\,.
\label{eq:deteq}
\end{equation}

For the first class of theory \eqref{action:NMMG-re}, the coefficients $\cal A$, $\cal B$, and $\cal D$ are given by
\begin{align}
\mathcal{A} = & 
\frac{3\,(\rho_\chi+P_\chi)}{m^2\mpl^2\,r^2}\Bigg[\frac{J\,\xi\,r}{\left(H+\frac{\dot{G}}{2\,G}\right)^2}\left(\frac{\kappa}{a^2}+\frac{m^2\,J\,\xi}{2\,G} \right) 
-\frac{J\,\xi}{\left(H+\frac{\dot{G}}{2\,G}\right)}\left(H-\frac{3\,\dot{G}}{2\,G}+\frac{\dot{J}}{J}+\frac{\dot{\xi}}{\xi}\right) +2\,(\Gamma-J\,\xi)\Bigg]
\nonumber\\
&+\frac{3}{(2\,G\,H+\dot{G})\,r^2}\left(\frac{\rho_\chi+P_\chi}{m^2\mpl^2}\right)^2\,\left(
\frac{m^2J\,\xi\,r}{H+\frac{\dot{G}}{2\,G}}+(1-3\,c_s^2)\dot{G}
\right)\,,\nonumber\\
%%%%
%%%%
\mathcal{B}= & \left(\frac{\rho_\chi+P_\chi}{m^2\mpl^2\left(H+\frac{\dot{G}}{2\,G}\right)\,r}\right)^2\left(
(1-3\,c_s^2)^2\frac{\dot{G}^2}{4\,G^2}-\frac{m^2r}{2\,G}(2\,\Gamma-3\,J\,\xi)
\right)\nonumber\\
&-\frac{J\,\xi\,(\rho_\chi+P_\chi)}{m^2\mpl^2\,r\,\left(H+\frac{\dot{G}}{2\,G}\right)^2}\vast\{
\frac{\ddot{\xi}}{\xi}
-\frac{1+r}{r}\left(H+\frac{\dot{\xi}}{\xi}\right)^2
\nonumber\\
&\qquad\qquad\qquad\qquad\qquad
+\left[
\frac{4\,\left(H+\frac{\dot{G}}{G}\right)\,[J\,\xi\,(r-1)+(r+1)\Gamma]}{J\,r\,\xi}
-\frac{\dot{G}\,[J\,\xi\,(7r-10)+4\,(r+1)\Gamma]}{2\,G\,J\,r\,\xi}
-\frac{2\,\dot{J}}{J\,r} - \frac{\dot{r}}{r}
\right]\left(H+\frac{\dot{\xi}}{\xi}\right)
\vast\}
\nonumber\\
&+\frac{(\rho_\chi+P_\chi)\,\xi}{4\,m^2\mpl^2\,J\,\left(H+\frac{\dot{G}}{2\,G}\right)^2}
\vast\{
\frac{4\,\left(H+\frac{\dot{G}}{2\,G}\right)^2}{r^2\,\xi^2}
\left[4\,(1+r)(\Gamma-J\,\xi)^2+J\,r\,\xi(10\,\Gamma-7\,J\,\xi)
\right]
\nonumber\\
&\qquad\qquad\qquad\qquad\qquad\qquad
+\frac{4\,J}{r^2\xi}\left(H+\frac{\dot{G}}{2\,G}\right)
\left[-2(\Gamma-J\,\xi)\left(\dot{r}+\frac{2\,(1+r)\,\dot{J}}{J}-\frac{(3+r)\dot{G}}{G}\right)
+2\,r(\dot{\Gamma}-\dot{J}\xi)
\right.
\nonumber\\
&\qquad\qquad\qquad\qquad\qquad\qquad\qquad\qquad\qquad\qquad\qquad\left.
+\frac{J\,r\,\xi}{2}\left(\frac{7\,\dot{G}}{G}-\frac{8\,\dot{J}}{J}\right)
\right]\nonumber\\
&\qquad\qquad\qquad\qquad\qquad\qquad+\frac{4\,J^2}{r}\left(\frac{2\,\ddot{G}}{G}-\frac{\ddot{J}}{J}\right)
+\frac{2\,J\,(r-6)\,\dot{J}\,\dot{G}}{G\,r^2}-\frac{2\,J^2\,\dot{r}}{r^2}\,\left(\frac{3\,\dot{G}}{G}-\frac{2\,\dot{J}}{J}\right)
+\frac{J^2(9-10\,r)\dot{G}^2}{G^2r^2}
\nonumber\\
&\qquad\qquad\qquad\qquad\qquad\qquad
+\frac{4(1+r)\,\dot{J}^2}{r^2}+\frac{4\,\kappa\,J}{a^2r\,\xi}[2\,(r-1)\Gamma+3\,J\,\xi]+\frac{4\,m^2J^2(3\,r-1)}{G\,r}(\Gamma-J\,\xi)+\frac{2\,m^2J^3\xi(2\,r+1)}{G\,r}
\vast\}\,,\nonumber\\
%%%%
%%%%
\mathcal{D}= & 
\left(
\frac{(1-3\,c_s^2)\,\dot{G}\,(\rho_\chi+P_\chi)}{m^2\mpl^2\,r\,(2\,G\,H+\dot{G})}\right)^2 \,.
\end{align}

For the second class, i.e. the projected theory \eqref{actionC:projected}, these coefficients are given by
\begin{align}
\mathcal{A} \equiv & \,
9c_s^2 (\rho_g+p_g)(\rho_\chi + p_\chi)(\dot{G}+2 G H) 
\Bigg[
\Big(	9c_s^2 \rho_\chi + 3(3c_s^2 -4)p_\chi + \rho_g -3 p_g\Big) \dot{G}
+2 G \Big(3 \dot{p}_g + H(\rho_g+9p_g)\Big)
\Bigg]
\,,\nonumber\\
\mathcal{B}\equiv & 
-9 c_s^2(1-3c_s^2)^2  (\rho_\chi + p_\chi)^2 (\rho_\chi + p_\chi + \rho_g +p_g)\dot{G}^2
-\frac{{\cal A}^2}{81 c_s^2 (\rho_g + p_g)^2(\rho_\chi + p_\chi)(\dot{G}+2GH)^2}
\nonumber \\
&
+\Bigg[\frac{ (13 \rho_g + 21 p_g)(\dot{G} + 2GH)+ 6(3c_s^2 -1)(\rho_\chi + p_\chi)\dot{G}}{9(\rho_g+p_g)(\dot{G} + 2GH)} 
-\frac{2}{3} \left(2-\frac{\mpl^2 M_{\rm GW}^2}{\rho_g+p_g}\right)
\Bigg]{\cal A}
\,,	\nonumber \\
\mathcal{D} \equiv &	
-(1-3c_s^2)^2 (\rho_\chi + p_\chi)^2 \dot{G}^2 
\Big[
(9c_s^2-1)\rho_g + 9(c_s^2-1)p_g 
-6  \mpl^2 M_{\rm GW}^2
\Big] \,.
\end{align} 

The solutions of Eq.~\eqref{eq:deteq} can then be written as
\begin{equation}
C^2 = \frac{1}{2}\,\left[c_s^2+\frac{\mathcal{B}}{\mathcal{A}}\pm\left(c_s^2-\frac{\mathcal{B}}{\mathcal{A}}\right)
\sqrt{
	1+4c_s^2 \frac{{\cal D}}{\mathcal{A}}\left(
	c_s^2 - \frac{\cal B}{\cal A}
	\right)^{-2} }
\right]\,.
\end{equation}
The conditions for avoiding gradient instability simply require a real sound speed, i.e. $C^2>0$ for both roots.
Due to the presence of the non-minimal coupling, it is not straightforward to distinguish between the matter perturbation and the scalar graviton. However, we observe that in the case $c_s^2=1/3$, the sound speeds in both theories become relatively simple with
\begin{equation}
C^2_1 \Big\vert_{c_s^2=1/3}= \frac{1}{3}\,,\qquad
C^2_2 \Big\vert_{c_s^2=1/3}= \frac{\mathcal{B}}{\mathcal{A}}\,.
\label{eq:NMLsound-speeds}
\end{equation}

\bibliography{transnoninvariant}

\end{document}